\documentclass[12pt,preprint]{emulateapj}

\def\be{\begin{equation}}
\def\ee{\end{equation}}
\def\bea{\begin{eqnarray}}
\def\eea{\end{eqnarray}}
\usepackage{graphicx}
\usepackage{amsmath}
\usepackage[colorlinks,
            linkcolor=blue,
            anchorcolor=blue,
            citecolor=blue]{hyperref}

\usepackage{color,txfonts}
\begin{document}

%\title{Fermi Large Area Telescope detection of a break in the $\gamma$-ray spectrum of the supernova remnant Puppis A}
\title{The SNR Puppis A Revisited with Seven Years of Fermi Large Area Telescope Observations}

\author{Yu-Liang Xin\altaffilmark{1,2}, Xiao-Lei Guo\altaffilmark{1,3}, Neng-Hui Liao\altaffilmark{1,4}
Qiang Yuan\altaffilmark{1,4}, Si-Ming Liu\altaffilmark{1,4}, Da-Ming Wei\altaffilmark{1,4}}
\email{E-mail: yuanq@pmo.ac.cn (QY); liusm@pmo.ac.cn (SML); dmwei@pmo.ac.cn (DMW)}
\altaffiltext{1}{Key laboratory of Dark Matter and Space Astronomy, Purple Mountain Observatory, Chinese Academy of Sciences, Nanjing 210008, China;}
\altaffiltext{2}{University of Chinese Academy of Sciences, Yuquan Road 19, Beijing, 100049, China;}
\altaffiltext{3}{Department of Physics and Institute of Theoretical Physics, Nanjing Normal University, Nanjing 210046, China}
\altaffiltext{4}{School of Astronomy and Space Science, University of Science and Technology of China, Hefei, Anhui 230026, China}

\begin{abstract}
Puppis A is a very famous and extensively studied supernova remnant (SNR) that shows strong 
evidence of shock-cloud interaction. We re-analyze the GeV $\gamma$-ray emission of it using 
seven years Pass 8 data recorded by the Fermi Large Area Telescope (Fermi-LAT). 
The morphology of the $\gamma$-ray emission is more compatible with that of the thermal X-ray and IR
emissions than the radio image, which suggests a possible correlation between the gamma-ray emitting 
region and dense clouds. The $\gamma$-ray spectrum in the energy range of 1-500 GeV shows a break 
at $7.92\pm1.91$ GeV with the photon indices of $1.81\pm0.08$ below the break and $2.53\pm0.12$ 
above the break, which can naturally explain the lack of TeV $\gamma$-ray emission from Puppis A. 
The multi-wavelength observations favor a hadronic origin for the $\gamma$-ray emission.
%Given the multi-wavelength data, only the hadron-dominated model seems to be plausible to describe
%the $\gamma$-ray emission and the radio break detected by the {\em WMAP} and {\em Planck} satellites.

\end{abstract}


\setlength{\parindent}{.25in}

\section{Introduction}

Supernova remnants (SNRs) are widely believed to be the most probable 
acceleration sites of Galactic cosmic rays (CRs) below energies of the 
knee \citep[see][for a review]{Hillas2005}. The high energy $\gamma$-ray
emission can pinpoint the presence of energetic electrons or nuclei.
SNRs interacting with dense molecular clouds (MCs) are expected to be
a class of the brightest sources in the $\gamma$-ray band. Indeed, the $\gamma$-ray
emission from several sources of this class has been detected by the
Fermi Large Area Telescope (Fermi-LAT), including IC443 \citep{Abdo2010a,Ackermann2013}, 
W44 \citep{Abdo2010b,Ackermann2013}, W28 \citep{Abdo2010c,Hanabata2014}, 
W51C \citep{Abdo2009},W49B \citep{Abdo2010d}, W30 \citep{Ajello2012}, 
Tycho \citep{Zhang2013}, Kesteven 27 \citep{Xing2015}, and Kesteven 41 \citep{Liu2015}.
The intense GeV $\gamma$-ray emission from these SNRs is generally believed
to be from the decay of neutral pions produced in inelastic collisions between accelerated protons
and the dense gas in MCs. Especially, $\gamma$-ray spectra of IC443, W44 and W51C have shown
spectral feature of $\pi^0$ decay\footnote{Note that the peak of the the $\pi^0$ decay spectrum 
is at half the $\pi^0$ mass, 67.5 MeV, so is not properly covered by present experiments to date. 
Hence the $\pi^0$ signature is of limited significance up to now, but should be addressable with 
future Fermi-LAT data or future experiments such as e-ASTROGAM \citep{Tatischeff2016} 
which extend to lower energies as required to detect the peak.} \citep{AGILE2011,Ackermann2013,Jogler2016}, 
which are considered the most direct evidence for the presence of 
relativistic nuclei acceleration in SNRs.
%\footnote{The hard sub-GeV spectrum of W44 measured 
%by AGILE was also interpreted due to $\pi^0$-decay origin \citep{AGILE2011}.}.

In addition, the $\gamma$-ray spectra of most SNRs mentioned above exhibit
remarkable spectral breaks in the 1-20 GeV band, steepening above the break 
energy. The spectral break can be explained by the energy-dependent diffusion of 
accelerated particles from the SNR shell into nearby molecular clouds \citep{
Aharonian1996,Gabici2009,Ohira2011,Li2010,Li2012}, the re-acceleration in crushed 
clouds \citep{Uchiyama2010} or the Alfv\'en wave evanescence in weakly ionized dense 
gas \citep{Malkov2011}.

Puppis A (G260.4-3.4) is a famous SNR with evident interaction with MCs. 
Its distance was determined to be $2.2\pm0.3$ kpc based on the observation 
of neutral hydrogen \citep{Reynoso2003}. At such a distance the diameter of 
Puppis A is estimated to be about 30 pc.
%CCO
A central compact object (CCO), RX J0822-4300, is located near the geometric center of 
Puppis A, which has been identified as the stellar remnant left after the supernova (SN) 
explosion \citep{Petre1996,Zavlin1999}. Based on observations of the proper motion of the 
CCO and optical filaments, the age of Puppis A was estimated to be $(4450\pm750)$ yr \citep{Becker2012,Winkler1988}, 
implying that this SNR is in the Sedov-Taylor evolutionary phase.
%BEK
There are two bright knots inside Puppis A, including the ``bright eastern knot''
and the ``bright northern knot'' \citep{Petre1982}. The X-ray spectral studies revealed
correlation between these knots and the shock-cloud interaction, making Puppis A to be 
the first X-ray-identified example of shock-cloud interaction in a relatively late phase 
of evolution \citep{Hwang2005,Katsuda2010,Katsuda2012}. No dense molecular gas is adjacent 
to the eastern knot, implying that the molecular clumps have been completely engulfed and 
destroyed by the shock front \citep{Paron2008}. The shock velocity of Puppis A is 
$\sim$ (600-1200) ${\rm km}$~s$^{-1}$, which was derived according to the oxygen and 
electron temperatures as well as the ionization timescale in the ejecta knot \citep{Katsuda2013}.

%X-ray/IR/VLA:
Puppis A is one of the brightest SNRs in the X-ray band. Each area of it has been observed by
several X-ray telescopes in orbit, e.g., {\em Einstein} \citep{Petre1982}, {\em ROSAT} \citep{Aschenbach1993},
{\em Suzaku} \citep{Hwang2008}, {\em Chandra} \citep{Hwang2005,Dubner2013}, and {\em  XMM-Newton \rm}
\citep{Hui2006,Katsuda2010,Katsuda2012,Dubner2013}. The X-ray emission detected is completely
thermal in origin, which mostly comes from the shocked interstellar medium \citep[ISM;][]{Hwang2005} 
except that there are a few isolated O-Ne-Mg-rich features associated with the SN ejecta
\citep{Hwang2005,Katsuda2008,Katsuda2010,Katsuda2013}. Recently, \citet{Dubner2013} showed the 
most complete and detailed X-ray view of Puppis A and confirmed that the SNR evolves in an 
inhomogeneous, probably knotty ISM. The hard component of the X-ray emission is spatially coincident 
with the distribution of neutral hydrogen column densities, 
implying the absorption of soft photons by neutral hydrogens \citep{Dubner2013,Reynoso2003}. 
The infrared (IR) images, observed by the {\em Spizer} Space Telescope, revealed an extremely good 
correlation with the X-ray emission, demonstrating that the thermal IR emission arises from 
swept-up interstellar dust, collisionally heated by the hot shocked gas \citep{Arendt2010}.

%GeV / TeV:
\citet{Hewitt2012} reported the detection of GeV $\gamma$-ray emission from Puppis A 
with the Fermi-LAT. With a luminosity of only $2.7\times 10^{34}\, (d/2.2\ {\rm kpc})^2$ erg~s$^{-1}$
between 1 and 100 GeV, Puppis A is among the faintest SNRs identified by the Fermi-LAT.
The morphology of the GeV $\gamma$-ray emission is spatially extended, which is compatible
with the X-ray morphology. The $\gamma$-ray emission is well described by a power law
spectrum with an index of 2.1. Considering the multi-wavelength data from the radio to $\gamma$-ray,
both leptonic and hadronic models are possible with different magnetic field strengths and 
different energies of relativistic particles. In addition, \cite{Hewitt2012} also 
reported a hint of a radio break or cutoff at $\sim40$ GHz using 7 years data from the {\em Wilkinson 
Microwave Anisotropy Probe} ({\em WMAP}). 

\citet{Abramowski2015} observed Puppis A in the very-high-energy 
(VHE; E $\geq$ 0.1 TeV) band with the High Energy Stereoscopic System (HESS). 
However, no significant VHE emission has been detected.
It has been suggested that the lack of the VHE signal implies that a spectral 
break or cutoff would occur at 280 or 450 GeV, assuming a power law with a simple 
exponential or a sub-exponential cutoff, respectively \citep{Abramowski2015}.

In this work, we report the detection of the GeV break in the spectrum of Puppis A, with 
the Pass 8 data recorded by the Fermi-LAT. In Section 2, the data analysis and results 
are presented, including the spatial and spectral analyses. The discussion about 
the origin of the non-thermal emission based on the multi-wavelength data is given 
in Section 3, followed by conclusions in Section 4.

\section{Data analysis}

\subsection{Data reduction}
We analysis the SNR Puppis A using the latest Pass 8 version of the Fermi-LAT data
taken in the period between August 4, 2008 (Mission Elapsed Time 239557418)
to August 4, 2015 (Mission Elapsed Time 460339204).  
The Galactic coordinate of Puppis A is (260.4, -3.4). 
It is close to the Galactic plane, but in a direction of the outer Galaxy. 
The surface brightness of the Galactic diffuse emission is smaller by a 
factor of $\sim$2 than that of Puppis A. A very bright gamma-ray source, 
the Vela pulsar, is about 3 degrees away from Puppis A, which actually 
affects the low energy analysis of Puppis A. Therefore, to prevent event 
contamination from the nearby Vela pulsar, only events with energies 
above 1 GeV are selected.
We select the ``source'' event class (evclass=128 \& evtype=3) and
exclude the data with zenith angle greater than $90^\circ$ to minimize contamination 
from the Earth limb. This analysis is performed within a square 
region of $14^\circ \times 14^\circ$ centered at the position of Puppis A
\citep[R.A.$=125.66^\circ$, Dec.$=-42.84^\circ$;][]{Acero2015},
which is referred to 
as a region of interest (ROI). We use the standard LAT analysis software, 
{\it ScienceTools} version {\tt v10r0p5}\footnote
{http://fermi.gsfc.nasa.gov/ssc/data/analysis/software/}, 
available from the Fermi Science Support Center, and adopt the instrumental 
response function (IRF) ``P8R2{\_}SOURCE{\_}V6''. The binned likelihood analysis 
method with {\tt gtlike} is used to fit the data. For the background subtraction,
the diffuse backgrounds including the Galactic emission and the isotropic
component, are modeled according to {\tt gll\_iem\_v06.fits} and
{\tt iso\_P8R2\_SOURCE\_V6\_v06.txt}\footnote
{http://fermi.gsfc.nasa.gov/ssc/data/access/lat/BackgroundModels.html}.
All sources in the third Fermi catalog \citep[3FGL;][]{Acero2015} within 
a radius of $15^\circ$ from the ROI center are included in the source model, 
which is generated with the user-contributed software 
{\tt make3FGLxml.py}\footnote{http://fermi.gsfc.nasa.gov/ssc/data/analysis/user/}.
Except that, two extra point sources, named as source A (R.A.$=129.707^\circ$, 
Dec.$=-44.083^\circ$) and source B (R.A.$=122.273^\circ$, Dec.$=-47.293^\circ$),
which are not in the 3FGL catalog are added to the model. During the fitting analysis, 
the normalizations and spectral parameters of all sources within a distance 
of $7^\circ$ from the ROI center, together with the normalizations of the two 
diffuse backgrounds, are left free.

\subsection{Spatial correlations}

\begin{figure*}[!htb]
\centering
\includegraphics[width=0.33\textwidth]{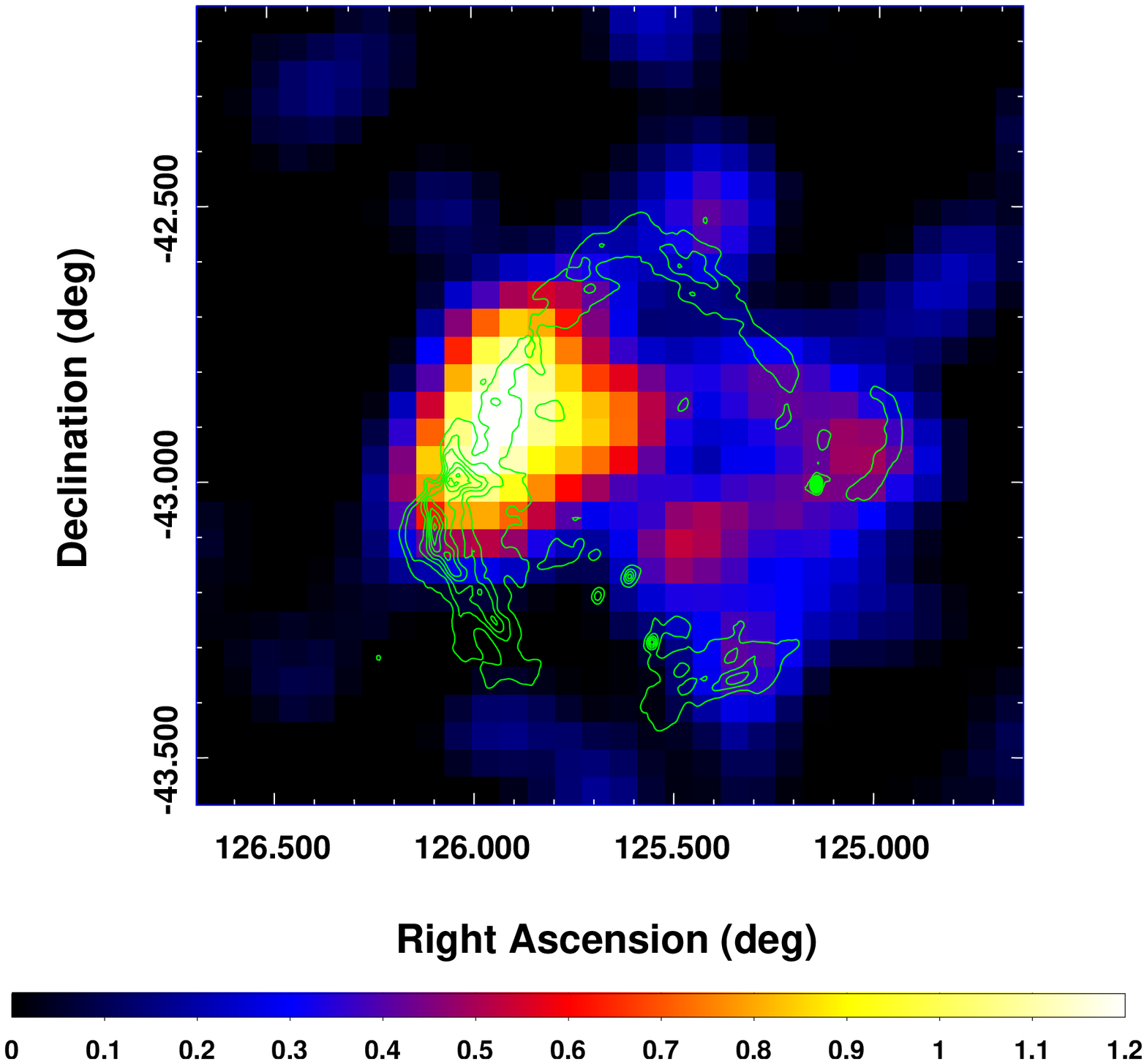}
\includegraphics[width=0.33\textwidth]{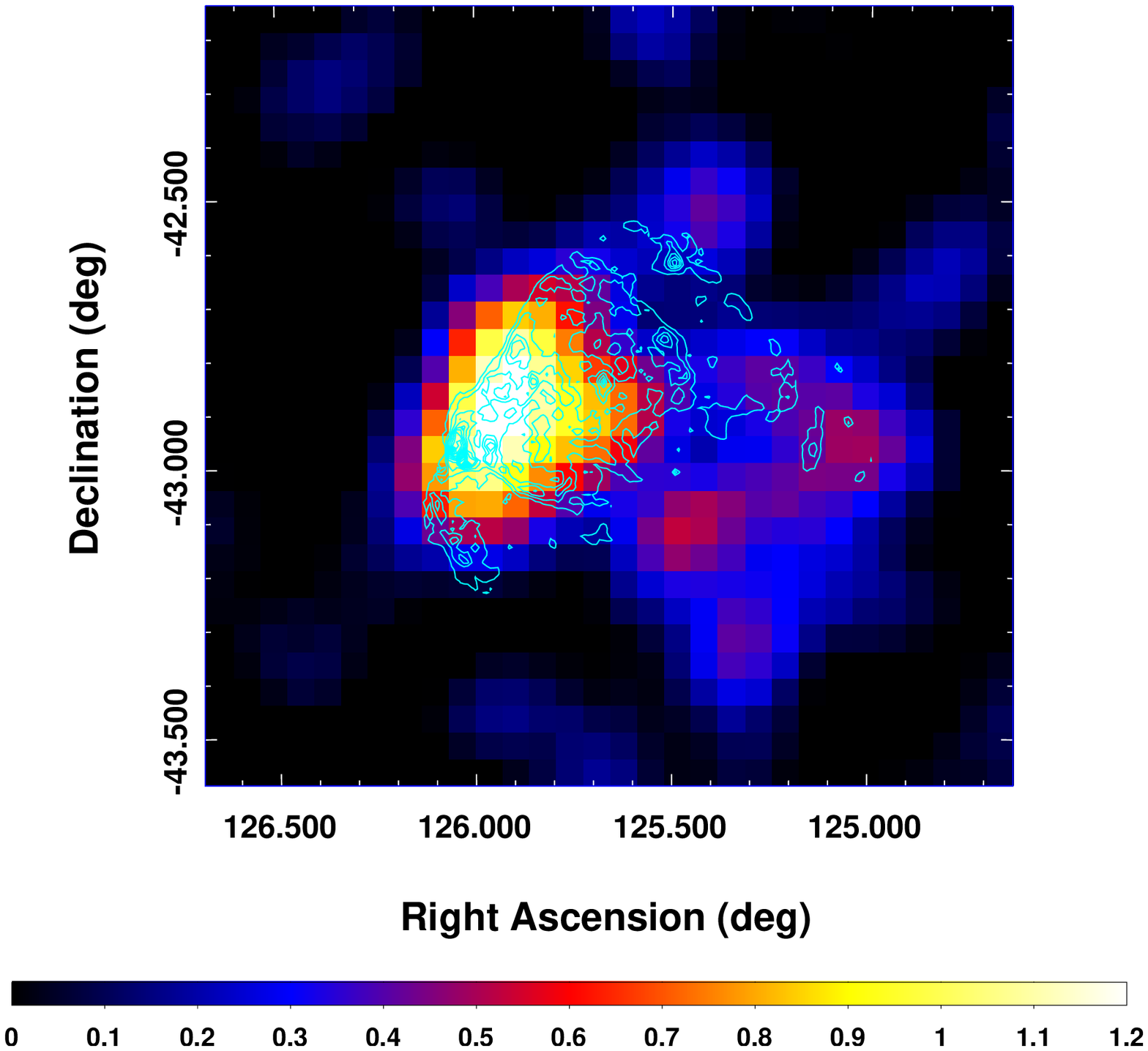}
\includegraphics[width=0.33\textwidth]{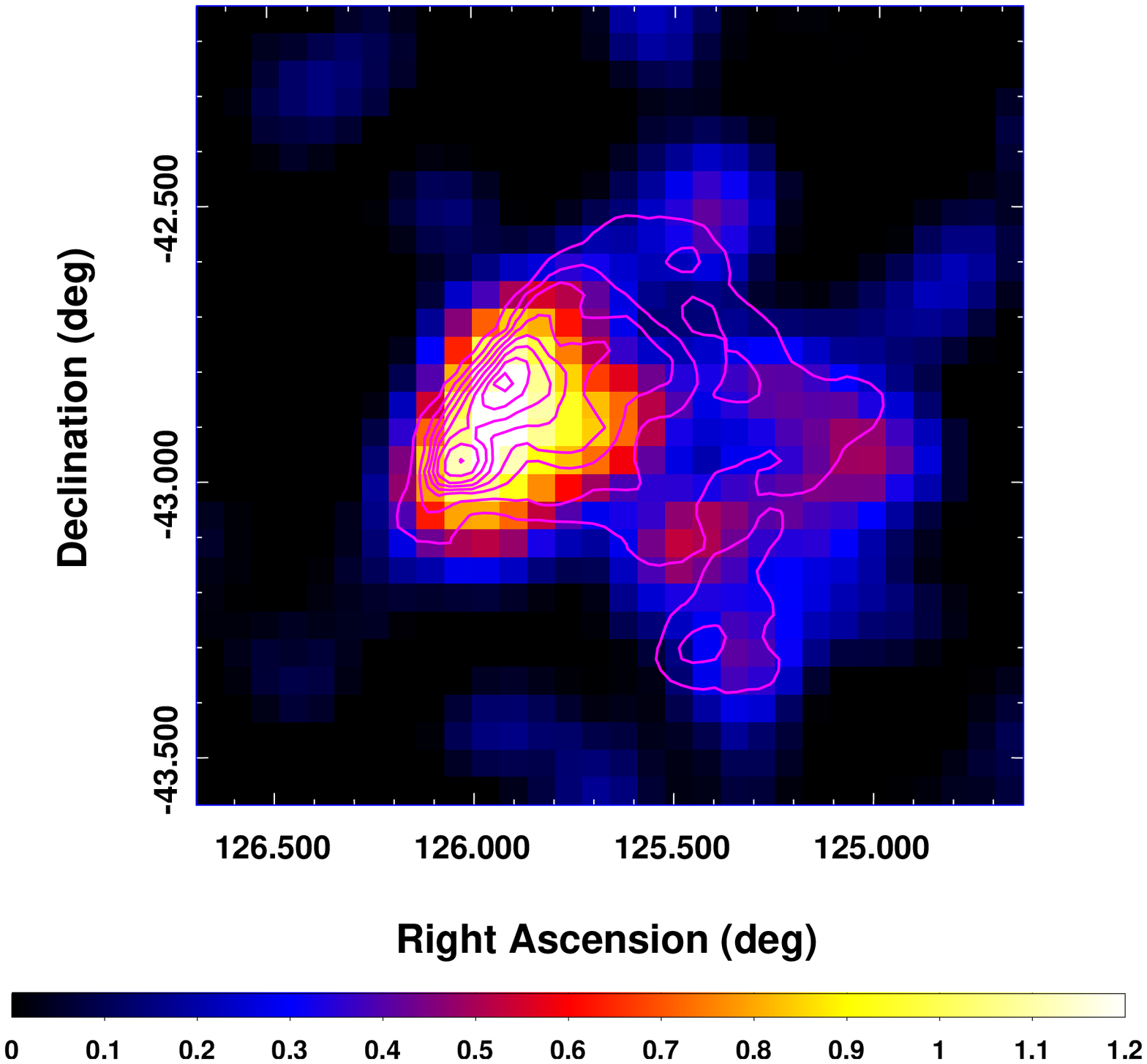}
\includegraphics[width=0.33\textwidth]{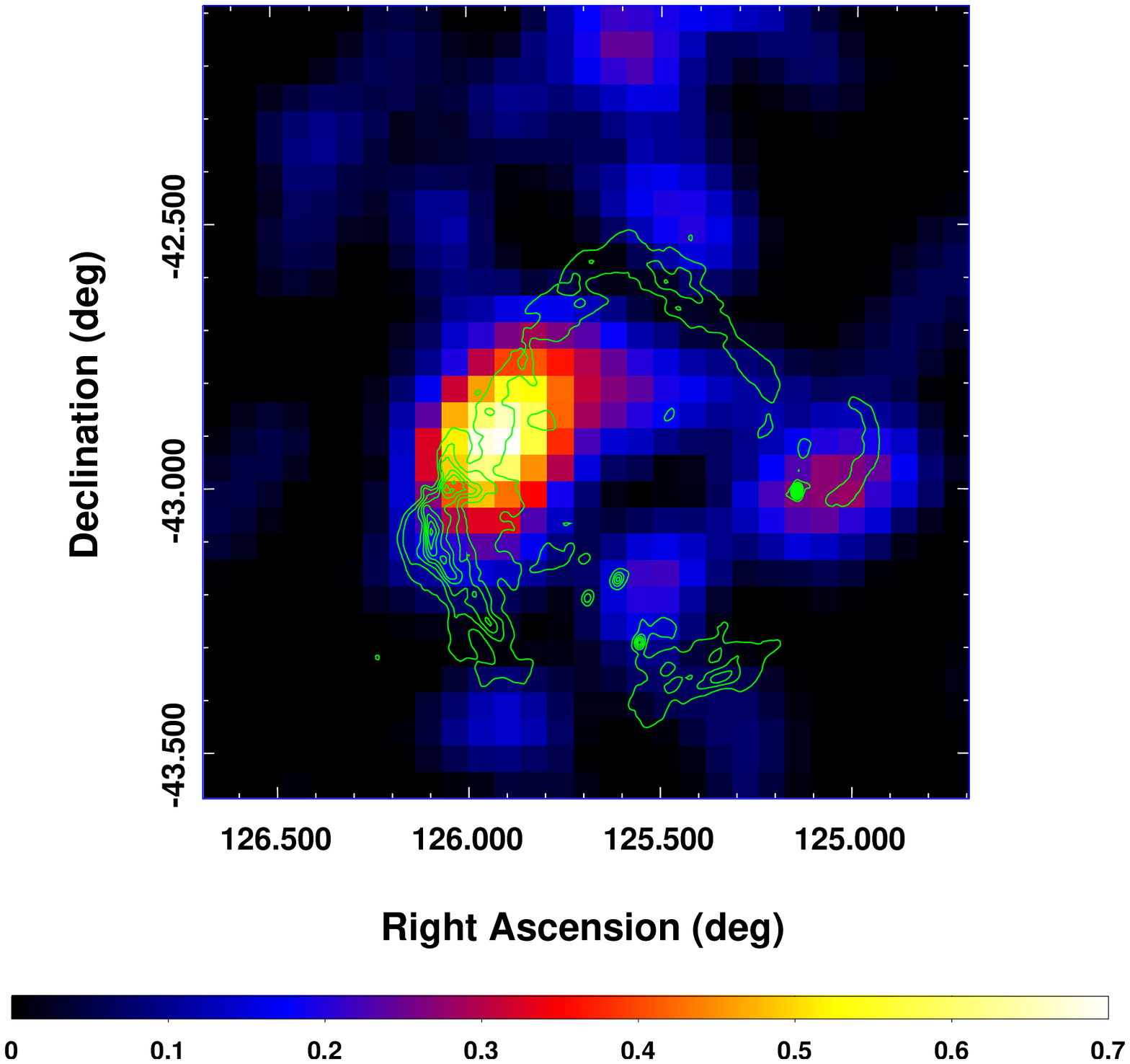}
\includegraphics[width=0.33\textwidth]{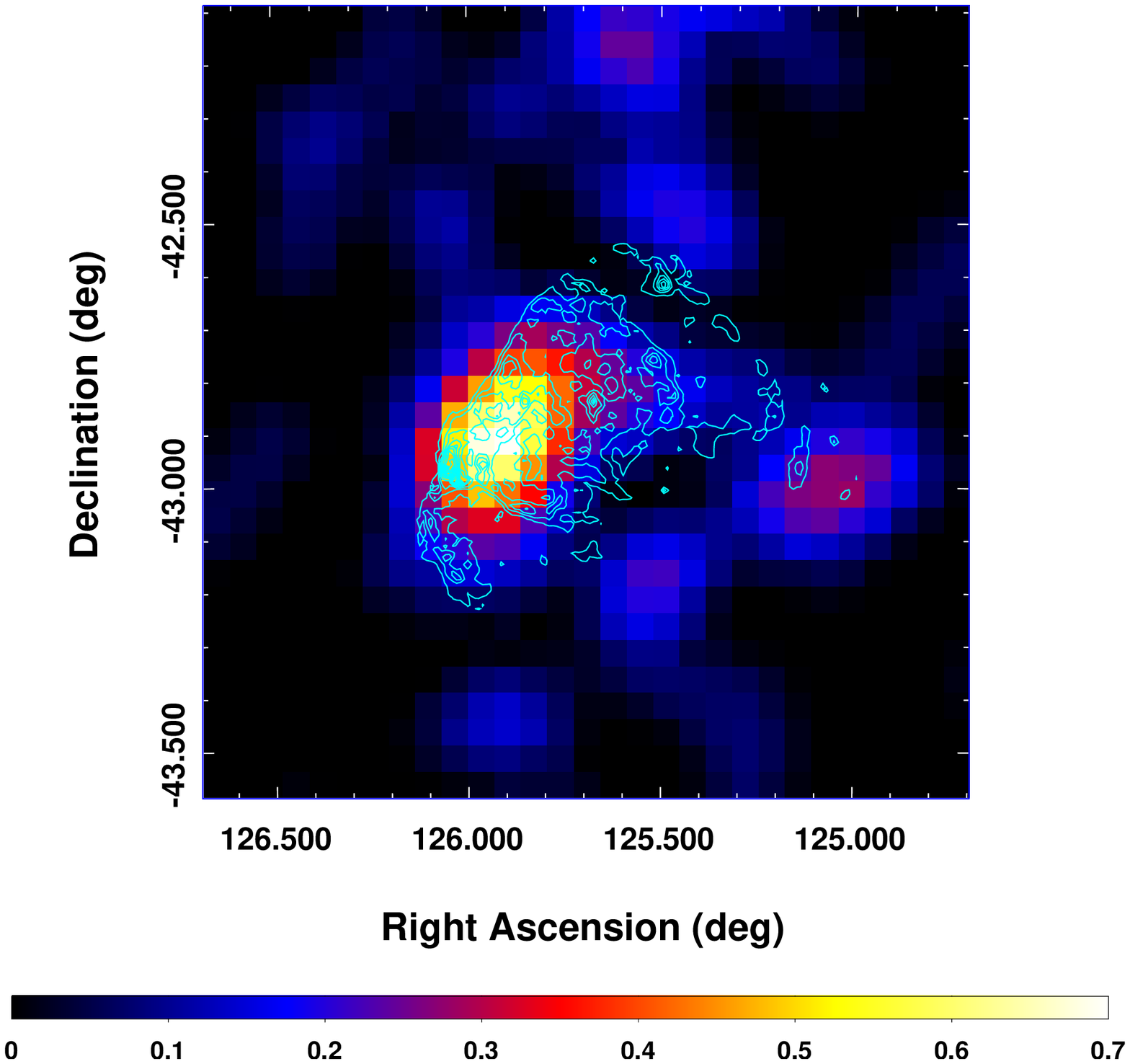}
\includegraphics[width=0.33\textwidth]{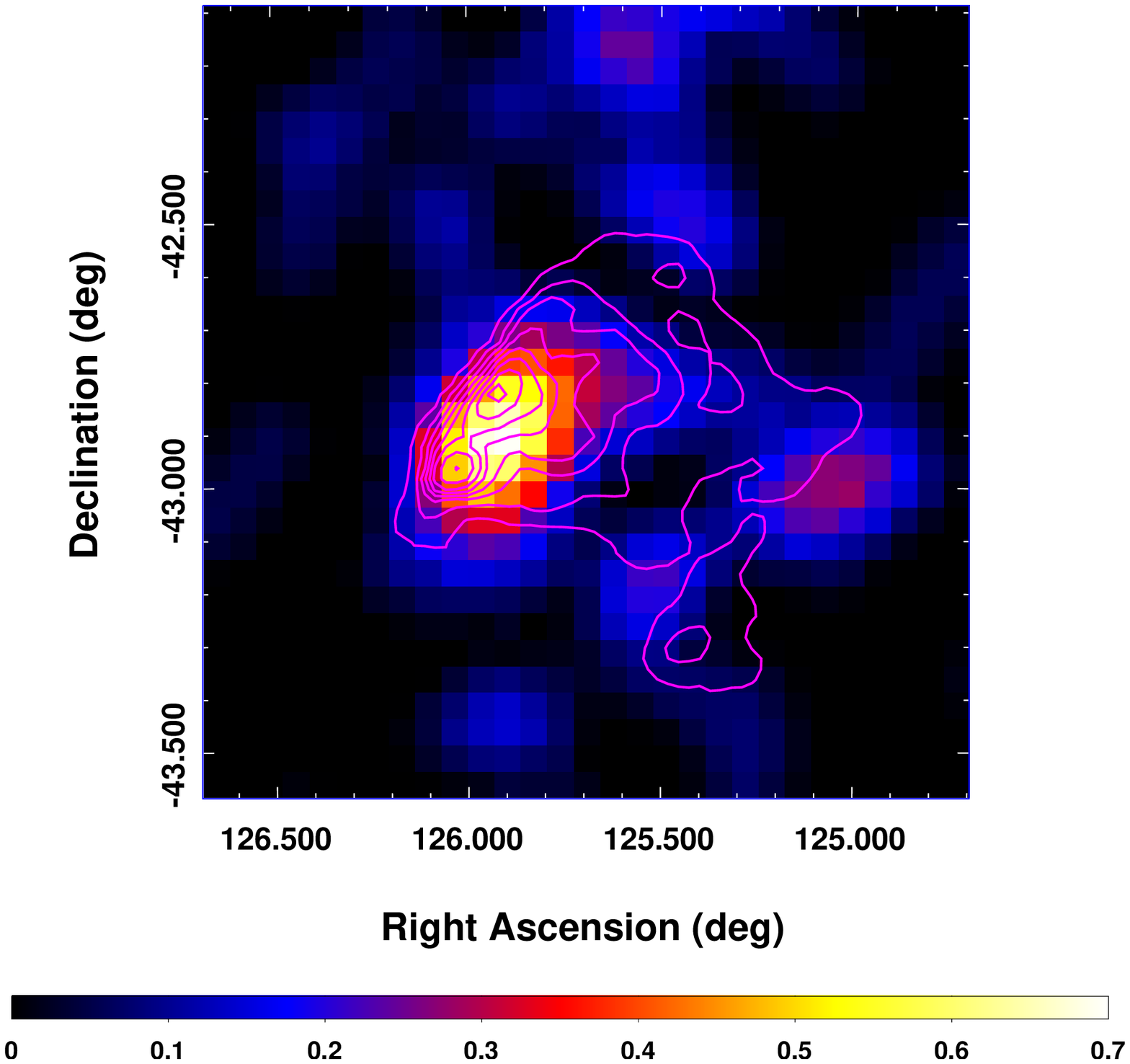}
\vspace{5em}
\caption{Residual counts maps of $1.5^{\circ} \times 1.5^{\circ}$ region centered at Puppis A
for photons above 5 GeV (PSF3-type data; top images) and 10 GeV (PSF3-type data; bottom 
images), derived by subtracting the best-fit model maps from the counts maps.
These maps are smoothed with a Gaussian kernel of $\sigma$ = $0.2^\circ$. 
Green contours (left) represent the radio image of Puppis A at 843 MHz from 
SUMSS \citep{Mauch2003}; cyan contours (middle) show the {\em ROSAT} HIR image 
of X-ray emission \citep{Petre1996} and magenta contours (right) display the IR 
image from IRAS satellite \citep{iras2005}.}
\label{fig:tsmap}
\end{figure*} 

The gamma-ray emission from Puppis A is extended. And an uniform disk with a radius 
of $0.37^{\circ}$ is suggested by the Fermi-LAT collaboration as the spatial template
for Puppis A \citep{Acero2015}.
In order to show the spatial correlation between the $\gamma$-ray and other energy 
bands, 
we created two residual counts maps by subtracting the best-fit model maps from 
the counts maps for photons above 5 GeV (top) and 10 GeV (bottom) in Fig. \ref{fig:tsmap} 
using the PSF3-type data (evclass=128 \& evtype=32), which has the best spatial resolution.
The radio contours at 843 MHz from the Sydney University Molonglo Sky 
Survey \citep[SUMSS;][]{Mauch2003}, the X-ray contours from {\em ROSAT} High 
Resolution Imager \citep[HRI;][]{Petre1996} and the infrared (IR) contours from 
IRAS satellite \citep{iras2005} are overplotted. 
The residual maps reveal that the GeV emission of Puppis A mainly 
concentrates in the northeast (NE) region and extends into the southwest (SW) region, 
which is more compatible with the X-ray/IR than the radio morphology.

Besides the uniform disk template, we also used three additional spatial 
templates according to multi-wavelength observations to fitting the $\gamma$-ray 
emission of Puppis A: the radio image at 843 MHz from SUMSS \citep{Mauch2003}, 
the X-ray image from {\em ROSAT} HRI \citep{Petre1996},
and the infrared (IR) image from IRAS satellite \citep{iras2005}.
The photon flux and TS value for each spatial template are list in Table \ref{table:spatial}.
As can be seen, the TS value for the X-ray image is larger than that for the uniform disk,
even though the spatial template of an uniform disk has more degrees of freedom. Comparing 
with the radio image, the X-ray/IR one is more consistent with the morphology of the $\gamma$-ray 
emission for the relatively higher TS value.

Hereafter, we adopt the X-ray image with the highest TS value as the spatial template for 
the whole SNR in the following spectral analysis.

\begin{table}[!htb]
\centering
%\scriptsize
\normalsize
%\tabletypesize{\small}
\caption {spatial distribution analysis for Puppis A between 1 GeV and 500 GeV}
\begin{tabular}{cccccc}
\hline \hline
Spatial  & Photon Flux                       & TS        & Degrees of \\
Template  & ($10^{-9}$ ph cm$^{-2}$ s$^{-1}$) & value      & Freedom\\
\hline
Uniform disk   & $8.23\pm0.29$ & 1825.9    &  5 \\
X-ray image    & $8.31\pm0.29$ & 1890.4    &  2 \\
Radio image    & $8.28\pm0.30$ & 1527.6    &  2 \\
Infrared image & $9.47\pm0.33$ & 1650.3    &  2 \\
\hline
\hline
\end{tabular}
\label{table:spatial}
\end{table}

\subsection{Spectral analysis}

\begin{figure}[!htb]
\centering
\includegraphics[height=2.3in]{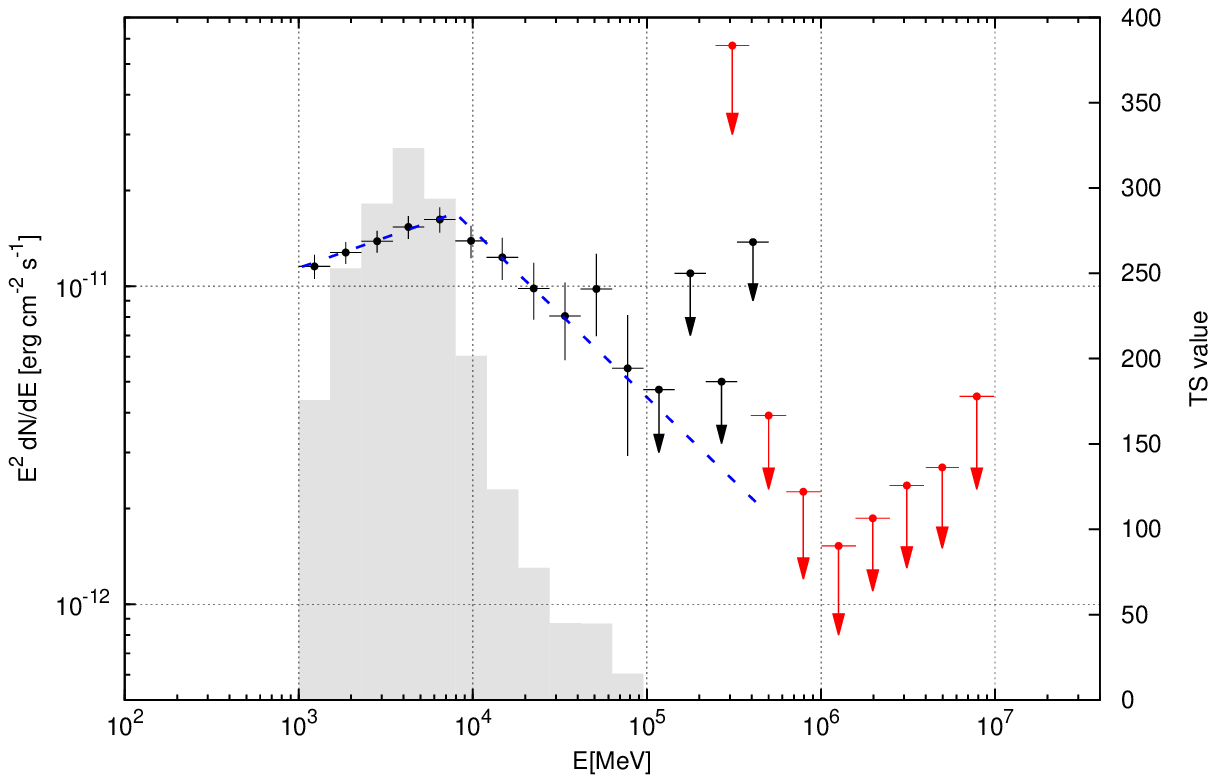} %
\includegraphics[height=2.3in]{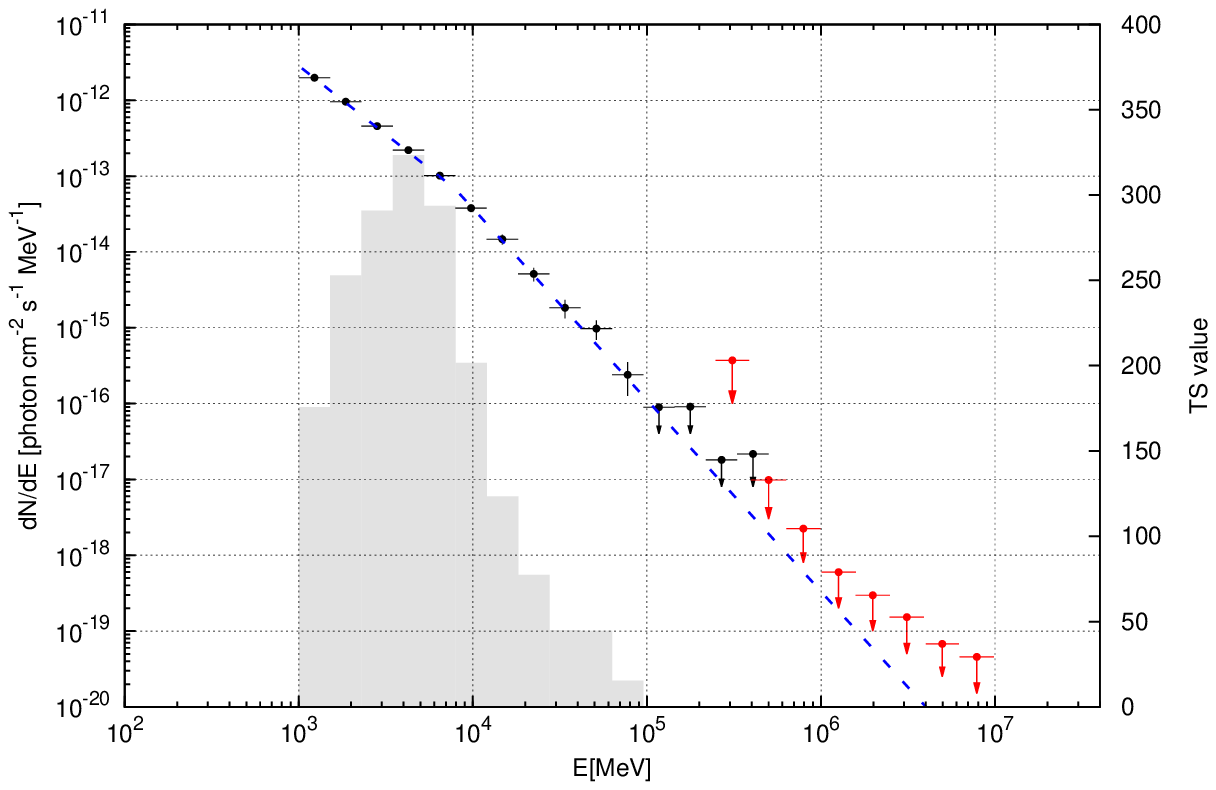} %
\hfill
\caption{The $E^2$ weighted energy spectrum (top) and photon spectrum (bottom) of the total SNR. 
The results of Fermi-LAT data are shown by black dots, with arrows indicating the $95\%$ upper limits. 
The gray histogram denotes the TS value for each energy bin. The upper limits in the TeV band (red) 
are from HESS observations \citep{Abramowski2015}. The best-fit BPL spectrum in the energy range 
from 1 GeV to 500 GeV is overplotted with the blue dashed line.}
\label{fig:totalsed}
\end{figure}

\begin{figure}[!htb]
\centering
\includegraphics[height=2.3in]{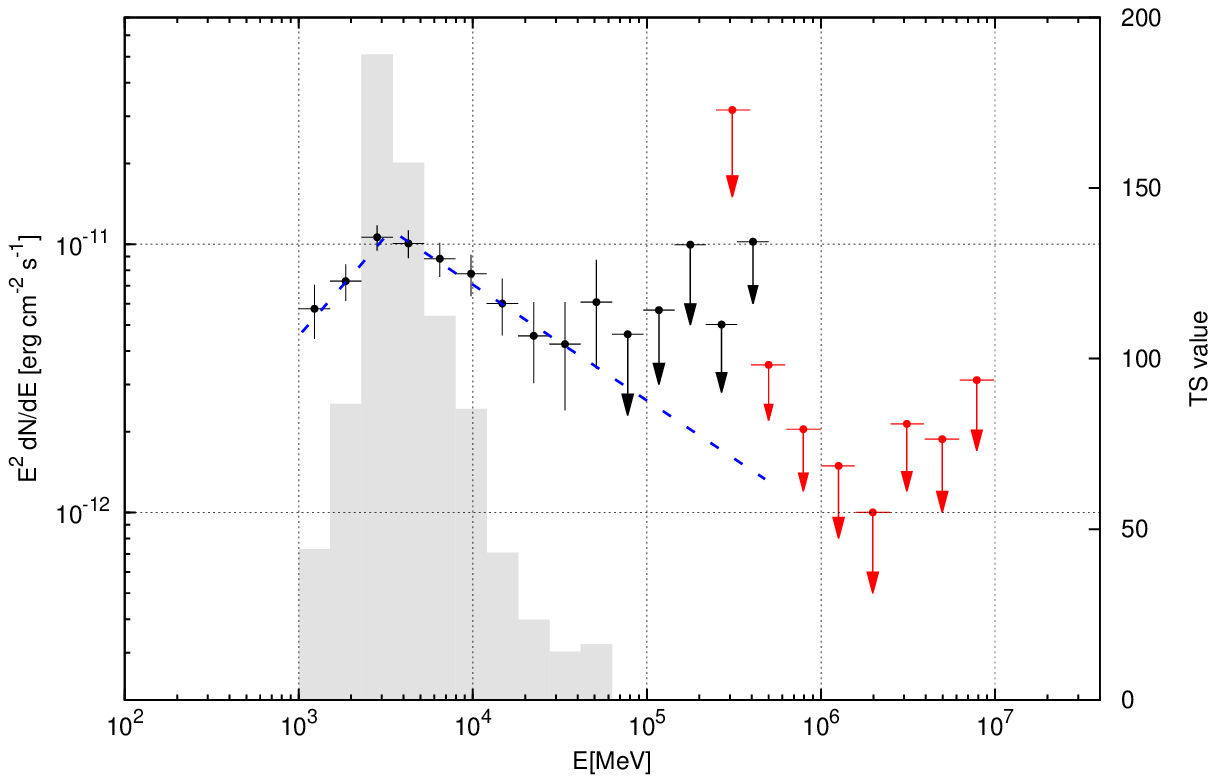} %
\includegraphics[height=2.3in]{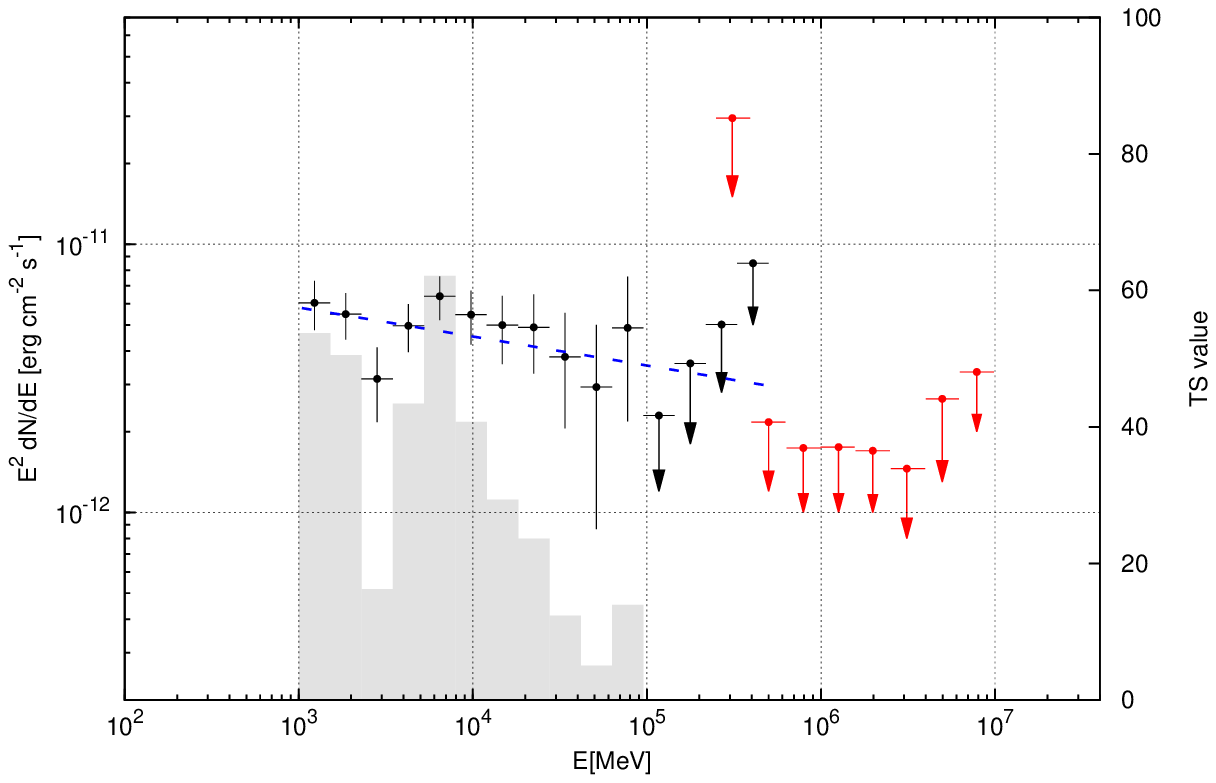} %
\hfill
\caption{Same as the top panel of Fig. \ref{fig:totalsed}, but for the eastern (top) and 
western (bottom) regions of the SNR separately. 
For the western region, a single PL spectrum is adopted to fit the Fermi-LAT data.}
\label{fig:partsed}
\end{figure}

We perform a spectral analysis of Puppis A in the energy range from 1 GeV to 
500 GeV with the spatial template of the X-ray image using the binned likelihood 
analysis method. First, we adopt a simple 
power law (PL) spectrum to fit the data. The spectral index
is found to be $2.07\pm0.03$, and the integral photon flux is $(8.31\pm0.29)\times10^{-9}$ 
photon cm$^{-2}$ s$^{-1}$, which is in agreement with the results given by \citet{Hewitt2012}.
We then test for a break in the spectrum of Puppis A using a broken power law (BPL) spectrum.
Significant improvement of the fitting can be found compared with the single PL model.
The results from the spectral fits are summarized in Table \ref{table:spectral}.
We tested for the significance of this spectral break using a likelihood ratio test:
${\rm TS_{break}} = -2\ln (L_{\rm PL}/L_{\rm BPL})$, where $L_{\rm PL}$ and $L_{\rm BPL}$
represent the likelihoods for the simple PL model and BPL model, respectively. 
We obtain ${\rm TS_{break}} = 35.4$, which corresponds to a significance of $\sim5.6\sigma$ 
for two additional degrees of freedom. Thus we conclude that the spectrum of Puppis A does 
have a break at $7.92\pm1.91$ GeV. The existence of the break and a relatively soft spectrum 
above the break energy, $\gamma_2=2.53$, can naturally explain the non-detection in the 
VHE $\gamma$-ray band \citep{Abramowski2015}.

Furthermore, to derive the $\gamma$-ray spectrum of Puppis A, we bin the data with 15 
logarithmically even energy bins between 1 GeV and 500 GeV, and perform the same likelihood 
fitting analysis to the data and the results are listed in Table \ref{table:seddata}.
The obtained energy spectrum weighted by $E^2$, overplotted with the best-fit BPL result, 
is shown in the top panel of Fig. \ref{fig:totalsed}. Note that the break of the spectrum should 
not be confused with the so-called $\pi^0$ peak of the photon spectrum, whose energy is much 
lower than that covered by our analysis (see the bottom panel of Fig. \ref{fig:totalsed} 
for an explicit illustration).
%{\bf The obtained energy spectrum is shown in the top panel of Fig. \ref{fig:totalsed},
%with the best-fitting BPL spectra between 1 GeV and 500 GeV overplotted.
%Note that the energy range does not cover the peak of the $\pi^0$ decay spectrum.
%Moreover, the photon spectrum of Puppis A is shown in the bottom panel of Fig. \ref{fig:totalsed}
%and the spectra with indices of 1.81 ($\gamma_1$) and 2.53 ($\gamma_2$) are overplotted, respectively.}

We also search for spectral difference between the eastern and the western regions of Puppis A, 
with a half-disk template each. The spectral data points of the eastern and the western regions 
are listed in Table \ref{table:seddata}. The spectra are shown in the 
top and the bottom panels of Fig. \ref{fig:partsed}, with the best-fitting spectra of each region 
overplotted. We find that the spectrum of the eastern region also shows a break with $E_b\sim3$ GeV, 
while for the western region a single PL is able to describe the GeV emission. 
However, considering the upper limit in the TeV band, a break or cutoff should be needed 
for the western region.

\begin{table}[!htb]
\centering
\scriptsize
\caption {Spectral Fit Parameters for PL and BPL between 1 GeV and 500 GeV with the spatial template of the X-ray image}
\begin{tabular}{cccccc}
\hline \hline
Spectral  & $\gamma_1$ & $\gamma_2$ & $E_b$ & Photon Flux                       & TS   \\
model     &            &            & (GeV)       & ($10^{-9}$ ph cm$^{-2}$ s$^{-1}$) & value \\
\hline
PL  & $2.07\pm0.03$ & $-$           & $-$           & $8.31\pm0.29$ & 1890.4   \\
BPL & $1.81\pm0.08$ & $2.53\pm0.12$ & $7.92\pm1.91$ & $8.01\pm0.29$ & 1925.8   \\
%PLE & $1.84\pm0.06$ & $-$ & $43.10\pm10.76$ & $7.98\pm0.29$ & 1857.4   \\
\hline
\hline
\end{tabular}
\label{table:spectral}
\end{table}

\begin{table*}
\centering
%\large
%\normalsize
\tabletypesize{\small}
\caption {Fermi-LAT spectral data points of Puppis A}
\begin{tabular}{cccccccccccccc}
\hline \hline
Energy band & photon flux of total  & TS value of & photon flux of eastern  & TS value of  & photon flux of western  & TS value of   \\
(GeV)          &SNR (ph cm$^{-2}$ s$^{-1}$) & total SNR  &region (ph cm$^{-2}$ s$^{-1}$) & eastern region & region (ph cm$^{-2}$ s$^{-1}$) &western region \\
\hline
1.00 -- 1.51  & ($2.44\pm0.21)\times 10^{-9}$ & 175.6 & ($1.22\pm0.28)\times 10^{-9}$ & 44.1 &($1.28\pm0.27)\times 10^{-9}$ & 53.7 \\
1.51 -- 2.29  & ($1.79\pm0.14)\times 10^{-9}$ & 252.9 & ($1.02\pm0.16)\times 10^{-9}$ & 86.7 &($7.68\pm1.51)\times 10^{-10}$ & 50.5 \\
2.29 -- 3.47  & ($1.28\pm0.10)\times 10^{-9}$ & 290.8 & ($9.82\pm1.06)\times 10^{-10}$ & 189.1 &($2.91\pm0.90)\times 10^{-10}$ & 16.2 \\
3.47 -- 5.25  & ($9.38\pm0.77)\times 10^{-10}$ & 323.3 & ($6.14\pm0.73)\times 10^{-10}$ & 157.4 &($3.03\pm0.61)\times 10^{-10}$ & 43.4 \\
5.25 -- 7.94  & ($6.54\pm0.60)\times 10^{-10}$ & 293.6 & ($3.56\pm0.52)\times 10^{-10}$ & 112.5 &($2.58\pm0.48)\times 10^{-10}$ & 62.1 \\
7.94 -- 12.01  & ($3.70\pm0.43)\times 10^{-10}$ & 201.5 & ($2.07\pm0.36)\times 10^{-10}$ & 85.2 &($1.46\pm0.33)\times 10^{-10}$ & 40.7 \\
12.01 -- 18.18  & ($2.17\pm0.33)\times 10^{-10}$ & 123.3 & ($1.06\pm0.25)\times 10^{-10}$ & 43.1 &($8.80\pm2.51)\times 10^{-11}$ & 29.3 \\
18.18 -- 27.51  & ($1.15\pm0.23)\times 10^{-10}$ & 77.3 & ($5.30\pm1.77)\times 10^{-11}$ & 23.5 &($5.70\pm1.87)\times 10^{-11}$ & 23.6 \\
27.51 -- 41.63  & ($6.20\pm1.70)\times 10^{-11}$ & 44.9 & ($3.26\pm1.42)\times 10^{-11}$ & 14.1 &($2.93\pm1.34)\times 10^{-11}$ & 12.3 \\
41.63 -- 63.00  & ($4.98\pm1.44)\times 10^{-11}$ & 44.7 & ($3.09\pm1.35)\times 10^{-11}$ & 16.3 &($1.49\pm1.05)\times 10^{-11}$ & 5.0\\
63.00 -- 95.33  & ($1.86\pm0.87)\times 10^{-11}$ & 15.3 &  $1.55\times 10^{-11}$ & $<5.0$  &($1.64\pm0.90)\times 10^{-11}$ & 13.9 \\
95.33 -- 144.27  & $1.05\times 10^{-11}$ & $<5.0$ & $1.26\times 10^{-11}$ & $<5.0$ & $5.10\times 10^{-12}$ & $<5.0$ \\
144.27 -- 218.32  & $1.61\times 10^{-11}$ & $<5.0$ & $1.46\times 10^{-11}$ & $<5.0$ & $5.27\times 10^{-12}$ & $<5.0$ \\
218.32 -- 330.39  & $4.86\times 10^{-12}$ & $<5.0$ & $4.86\times 10^{-12}$ & $<5.0$ & $4.86\times 10^{-12}$ & $<5.0$ \\
330.39 -- 500.00  & $8.82\times 10^{-12}$ & $<5.0$ & $6.54\times 10^{-12}$ & $<5.0$ & $5.44\times 10^{-12}$ & $<5.0$ \\
\hline
\hline
\end{tabular}
\label{table:seddata}
\tablecomments{For energy bins with TS values smaller than 5.0, 
the upper limits at 95\% confidence level are calculated.}
\end{table*}

\section{Discussion}

\subsection{Morphology and $\gamma$-ray spectrum}

From the above analysis, the GeV $\gamma$-ray emission is spatially well correlated 
with the morphology in the X-ray/IR rather than the radio band, which means that
there should be some dense clouds in the $\gamma$-ray emitting region \citep{Dubner2013,Abramowski2015}.
However, no CO emission was detected in such regions, which can be attributed to 
dissociation of molecules by the radiative precursor of the SNR due to the 
photoionization and photodissociation effects \citep{Paron2008}
or there is some CO-dark gas in the $\gamma$-ray 
emitting region which can not be traced by CO observations \citep{Grenier2005}.

There may be another explanation for the spatial correspondence between the 
$\gamma$-ray and the X-ray emissions. \citet{Dubner2013} shows that Puppis A 
displays a softening of the X-ray emission from NE to SW region, corresponding to 
the variation of column density of neutral hydrogen \citep{Reynoso2003,Hwang2005}. 
The higher-density column in the NE region may be responsible for the absorption of
soft X-ray photons resulting in the spatial distribution of the X-ray emission.
Therefore, there may exist some dense HI clouds in the vicinity of the central 
part of Puppis A. These HI clouds are illuminated by energetic particles from
Puppis A to emit strong $\gamma$-ray emissions.

Spatial correlation among the $\gamma$-ray and the thermal X-ray emission 
also exists in SNR W51C \citep{Abdo2009} and Cygnus Loop \citep{Katagiri2011}. 
However, for W51C, the $\gamma$-ray emission is also consistent with the radio 
extension. And for Cygnus Loop, the morphology of the X-ray emission is shell-like 
and the H$\alpha$ filaments together with the radio continuum emission in the 
northern part of Cygnus Loop are correlated with the $\gamma$-ray emission spatially, 
which are very different from Puppis A. \citet{Humensky2015} reported that the $\gamma$-ray 
emission of the SNR IC443 is anticorrelated with its thermal X-ray emission, which is 
opposite to that of Puppis A.

The spectral analysis of Puppis A reveals the existence of a break at $7.92\pm1.91$ GeV 
in the spectrum and the GeV break is usually linked to SNRs interacting with 
molecular clouds, like IC443, W44 and W51C, etc. These SNR/MC systems exhibit a 
spectral break in the 1-20 GeV band and they are typically brighter in the GeV band 
than in the TeV band \citep{Funk2015,Guo2017}, which is similar to Puppis A.
Therefore, the $\gamma$-ray radiation mechanism of Puppis A should be very similar 
to that of the SNR/MC systems. Compared with the other SNR/MC systems, the lower luminosity 
of Puppis A in the GeV band suggests that the gas density in Puppis A may not be as high as 
the others \citep{Hewitt2012}. 
The spatial variation of the GeV spectrum shows that the broken power-law spectrum of 
the overall emission may have a more complicated origin due to superposition of emission 
from different regions with different spectra. Nevertheless, these results suggest that the 
spectral break decreases with the decrease of shock speed as indicates by the east-west 
asymmetry of the remnant in agreement with the model proposed by \citet{Ohira2016} 
and the general trend of spectral evolution discovered by \citet{Zeng2017}. 

The $\gamma$-ray emission can be produced via three radiation mechanisms: Inverse Compton
Scattering (ICS) or bremsstrahlung process of high energy electrons, and the $\pi^0$ decay 
due to inelastic $pp$ collisions with the latter two processes proportional to density 
of the background plasma. The spatial correlation between the GeV $\gamma$-ray 
emission and the potential clouds traced by thermal 
X-ray emission makes the hadron-dominated model more preferable \citep{Katz2008}. In addition, 
the theoretical explanation of GeV break also favors the hadronic 
scenario \citep{Aharonian1996,Gabici2009,Ohira2011,Li2010,Li2012,Uchiyama2010,Malkov2011}.

\subsection{Spectral Energy Distribution}

In the radio band of Puppis A, \cite{Hewitt2012} reported a hint of a radio break or cutoff 
at $\sim40$ GHz using the {\em WMAP} data and such spectral structure was also detected 
by {\em Planck} satellite \citep{Planck2016}. The X-ray emission of Puppis A is completely
thermal in origin and no non-thermal component was detected. However, considering the 
fact that some SNRs as young as Puppis A, like RCW 86, still exist non-thermal X-ray 
emission \citep{Lemoine2012}, the non-thermal component of Puppis A should not be much 
lower than the thermal component. Therefore, the radio structure at $\sim40$ GHz is 
more like a break rather than a cutoff. Here, We discuss different radiation models 
in light of the multi-wavelength data, including the spectral structures of the radio 
and GeV break.

Considering the radio break, the spectra of electrons is assumed to be a smoothly broken 
power law with an exponential cutoff (SBPL) in the form of
\begin{eqnarray}
%$\dfrac{dN_e}{dE} \propto E^{-\alpha_L} \left(1 + \left(\dfrac{E}{E_{e,\rm br}}\right)^{2} \right)^{-(\alpha_H-\alpha_L)/2} \exp \left(- \dfrac{E}{E_{e,\rm cut}} \right)$.
\dfrac{dN_e}{dE} \propto E^{-\alpha_{e}} \left(1 + \left(\dfrac{E}{E_{e,br}}\right)^{2} \right)^{-\Delta \alpha_{e}/2} \exp \left(- \dfrac{E}{E_{e, \rm cut}} \right).
\label{eq:e_spectra}
\end{eqnarray}
And the spectra of protons is adopted to be a power law with an exponential cutoff (PL):
\begin{eqnarray}
\dfrac{dN_p}{dE} \propto E^{-\alpha_{p}} \exp \left(- \dfrac{E}{E_{p, \rm cut}} \right).
\label{eq:p_spectra}
\end{eqnarray}
where $\alpha_{i}$ and $E_{i, \rm cut}$ are the spectral index and the cutoff energy of particles, 
respectively, for $i = e$ or $p$. $E_{e,\rm br}$ is the break energy and $\Delta \alpha_{e}$ 
is the spectral variation of electrons.

In the modeling, we set the spectral index of protons $\alpha_{p}$ being equal to the
index of electrons below the break energy $\alpha_{e}$, assuming that this part of the 
spectra are determined by the acceleration process. In addition, we assume $\Delta \alpha_{e} = 1$ 
considering the energy-loss mechanism of electrons.
The distance of Puppis A is adopted to be 2.2 kpc \citep{Reynoso2003}, and the radius 
is $r\approx17.6$ pc for an angular size of $55'$ at such a distance. Besides the cosmic microwave 
background (CMB), two IR radiation fields are taken into account ($T_1=150$ K, $u_1=0.48$ eV cm$^{-3}$
and $T_2=45$ K, $u_2=0.2$ eV cm$^{-3}$), which are determined by the dust emission of
Puppis A \citep{Arendt1991, Hewitt2012}. The gas density, $n_{\rm gas}$, ranges from $0.5$ cm$^{-3}$ 
to $4.0$ cm$^{-3}$, which corresponds the variation of X-ray brightness from the 
northeast to southwest region of Puppis A \citep{Arendt2010,Hewitt2012}. 
\citet{Arendt1991} and \citet{Arendt2010} gave the dust mass of 0.25 $M_{\odot}$ for Puppis A
by infrared observations, and considering a typical dust-to-gas ratio of 0.0077 for ISM, the gas mass of 33 $M_{\odot}$ 
is derived \citep{Arendt1991}. However, the dust-to-gas ratio for the post-shock region of Puppis A is much lower than
that for ISM and the gas mass should be much larger \citep{Arendt2010}. 
Nevertheless, the value of the gas mass derived from this method should be the lower limit.  
Meanwhile, the total mass of the gas calculated by 
$M_{\rm gas}$ = $\dfrac{4}{3}$ $\pi$ $r^{3}$ $n_{\rm gas}$ $m_{\rm p} = 2.26\times10^{3} (n/4.0\,\mathrm{cm}^{-3}) M_{\odot}$, 
should be considered as an upper limit for the inhomogeneity of the actual gas density.

\begin{table*}
\centering
%\large
\normalsize
%\tabletypesize{\small}
\caption {Parameters for the models}
\begin{tabular}{cccccccccccccccc}
\hline \hline

Model & $\alpha_{p}$ & $\alpha_{e}$ & $\Delta \alpha_{e}$ & $E_{p, \rm br}$ & $E_{p, \rm cut}$ & $E_{e, \rm br}$ & $E_{e, \rm cut}$ & $B$ & $W_e^{\,\,\,\,\text{a}}$ & $n_{\rm gas}\times W_p^{\,\,\,\,\text{a,b}}$ &  $\varepsilon_p^{\,\,\,\,\text{c}}$ \\
      &              &              &       &     (TeV)       &     (TeV)       &      (GeV)      &      (TeV)   &    ($\mu$G)  &   ($10^{49} \mathrm{erg}$) & ($10^{49} \mathrm{erg}~cm^{-3}$) &  ($eV~cm^{-3}$)  \\
\tableline
ICS-dominated  & $1.80$ & $1.80$ & $1.0$  & $-$  & $10.0$ & $20.0$  & $0.8$  & $6.0$ & $1.7$ & $2.5$   &   $46.5$  \\
\hline
Brems-dominated  & $1.85$ & $1.85$ & $1.0$  & $-$   & $4.0$ & $18.0$  & $4.0$  & $11.0$ & $0.7$ & $1.4$   &   $3.3$ \\
\hline
Hadron-dominated (PL)  & $1.90$ & $1.90$ & $1.0$  & $-$  & $0.6$ & $6.0$   & $6.9$  & $72.0$ & $3.2\times10^{-2}$ & $30.0$   &   $69.8$ \\
\hline
Hadron-dominated (SBPL)  & $1.90$ & $1.90$ & $1.0$  & $0.2$  & $>6.9$ & $6.0$   & $6.9$  & $72.0$ & $3.2\times10^{-2}$ & $30.0$   &   $69.8$ \\
\hline
\hline
\end{tabular}
\label{table:model}
\tablecomments{
\\
a) The total energy of relativistic particles, $W_{e,p}$, is calculated for $E > 1$ GeV.\\
b) For the ICS-dominated model, the gas density, $n_{\rm gas}=0.5$ is adopted; for the Bres- and Hadron-dominated models, $n_{\rm gas}=4.0$ is adopted.\\
c) The energy density of CRs, $\varepsilon_p$, is calculated by $n_{\rm gas}\times W_p$ = $\dfrac{M_{gas}}{m_p}$ $\varepsilon_p$. Here the total mass of the gas, $M_{gas}$ is defined as $M_{\rm gas}$ = $\dfrac{4}{3}$ $\pi$ $r^{3}$ $n_{\rm gas}$ $m_{\rm p}$, assuming a homogeneous gas density for the whole volume of Puppis A.
}
\end{table*}

\begin{center}
\begin{figure*}[htbp]
\centering
\includegraphics[height=2.1in]{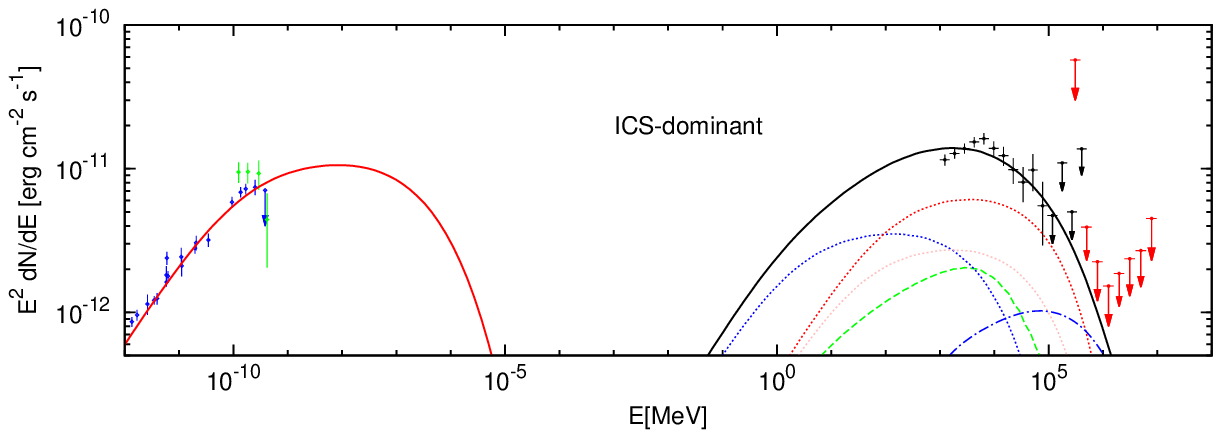} %IC n=0.5
\includegraphics[height=2.1in]{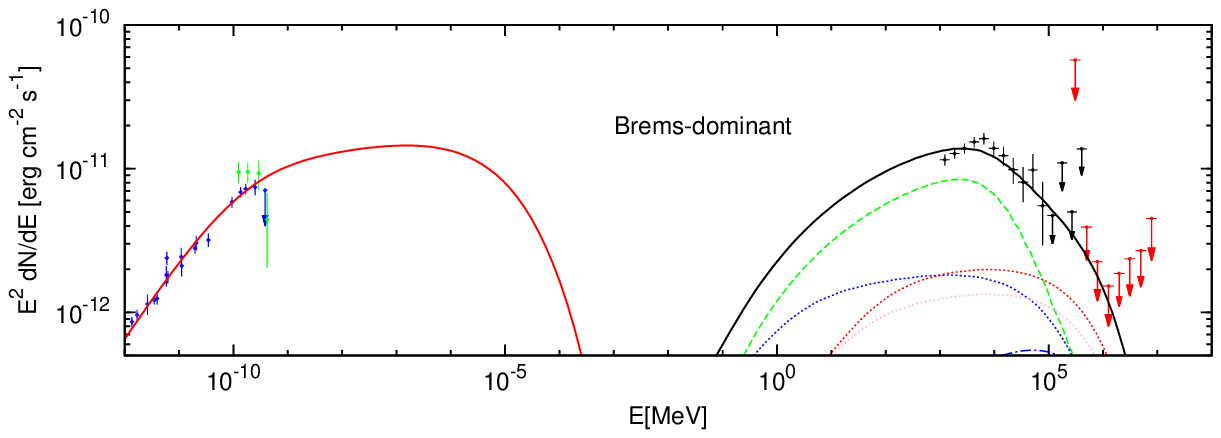} % Brems
\includegraphics[height=2.1in]{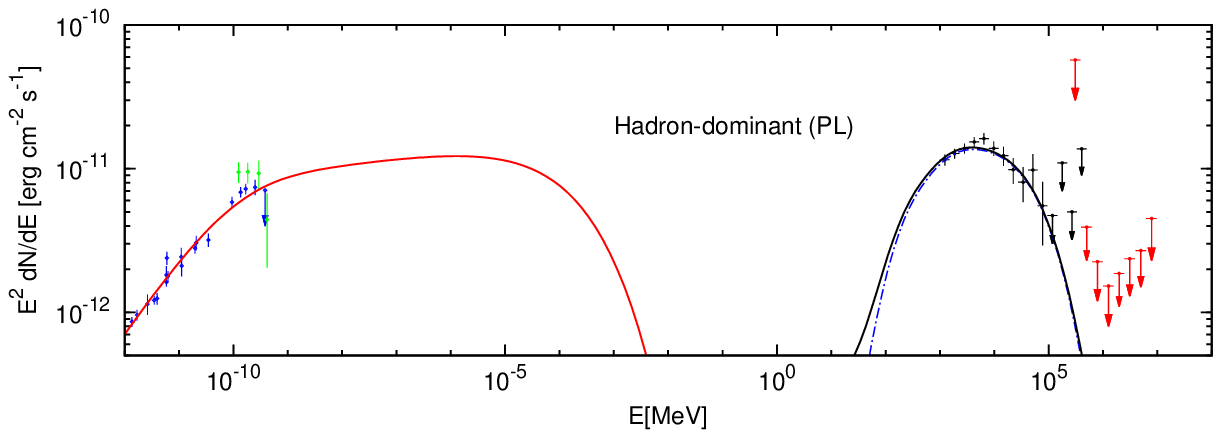} % pion  power law
\includegraphics[height=2.1in]{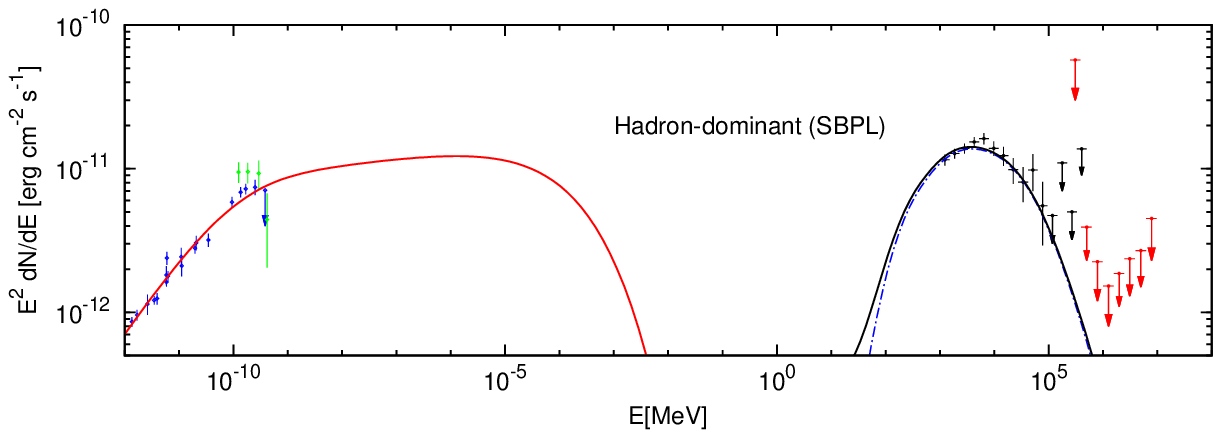} % pion smoothly broken power law
\caption{Different emission models fitting on the multi-wavelength data of Puppis A
with the parameters listed in Table \ref{table:model}. The hadron-dominated 
models with the spectra of a power law and a smoothly broken power law for protons are 
plotted separately. In each panel, the radio data marked by blue and green points are 
observed by {\em WMAP} \citep{Hewitt2012} and {\em Planck} satellites \citep{Planck2016}, 
respectively. And the radio emission is dominated by the synchrotron component, shown as 
the red solid curve. For the $\gamma$-ray emission, the contributions from ICS, bremsstrahlung, 
and $\pi^0$-decay processes are shown as the dotted, dashed and dotted-dashed lines, respectively.
The ICS emission includes three components from CMB (blue) and two infrared (red and pink)
photon fields. The black solid line represents the sum of different radiation components.}
\label{fig:multi-sed}
\end{figure*}
\end{center}

\subsection{ICS-dominated model}

%{\color {red} For the ICS-dominated model and Bremsstrahlung-dominated model, 
%is it necessary to discuss the two-zone model considering the radio and GeV morphologies 
%are not corresponding.}

The multi-wavelength spectrum energy distribution (SED) with the ICS-dominated model is shown in the top panel of
Fig. \ref{fig:multi-sed} and the model parameters are compiled in Table \ref{table:model}.
For the ICS-dominated model, the spectral index of electrons is 1.8.
The break and cutoff energies of electrons are 20 GeV and 0.8 TeV, respectively.
And the total energy of electrons above 1 GeV is estimated to be 
$W_{\rm e}\approx 1.7 \times 10^{49}~{\rm erg}$. A magnetic field strength of
about 6.0 $\mu$G is needed to explain the flux in the radio band. Here the gas 
density, $n_{\rm gas}$, is adopted to be $0.5$ cm$^{-3}$ to reduce the contribution 
from bremsstrahlung emission. The total energy of protons above 1 GeV, $W_{\rm p}$, 
is constrained to be less than $\sim 5.0 \times 10^{49} (n/0.5\,\mathrm{cm}^{-3})^{-1}\ \mathrm{erg}$,
considering the upper-limits at several hundreds of GeV. Here, the cutoff energy 
of protons, $E_{p, \rm cut}$, is adopted to be 10 TeV.

This model gives a reasonable fit to the overall spectrum. However, the inferred magnetic
field, $\sim$ 6 $\mu$G, is relatively low compared with that derived from an equipartition 
assumption \citep{Dubner2013}. In addition, the energy content of energetic electrons 
is too high if SNRs indeed dominate the cosmic ray fluxes observed at earth.

\subsection{Bremsstrahlung-dominated model}

To explain the emission from the radio to $\gamma$-ray band with the bremsstrahlung-dominated
model, we assume a gas density of $4.0$ cm$^{-3}$. It is interesting to note that a broken power-law 
electron distribution with an index of 1.85, a break energy of 18 GeV and a cutoff energy of 4.0 TeV 
can reproduce the spectral breaks in both the radio and $\gamma$-ray band. 
The corresponding magnetic field is 11.0 $\mu$G and the total energy of electrons above 1 GeV is 
$7.0 \times 10^{48}\mathrm{erg}$.
Here the total energy of protons above 1 GeV should be less than 
$\sim 3.5 \times 10^{48} (n/4.0\,\mathrm{cm}^{-3})^{-1}\ \mathrm{erg}$
for a typical cutoff energy of 10 TeV for protons to avoid overestimation of the VHE $\gamma$-ray
emission. The modelled SED is shown in the middle-upper panel of Fig. \ref{fig:multi-sed} and the
model parameters can be seen in Table \ref{table:model}.
The spectral fit is also reasonable and the model parameters are well constrained. 
However, this model implies a lower energy content in energetic protons than that in electrons, 
which is not consistent with the supernova origin of Galactic cosmic rays.

The above lepton-dominated models can marginally explain the $\gamma$-ray emission and the radio break, 
considering the statistic errors of the GeV data. However, for these two models, the 
estimated electron to proton ratio must be much larger than 0.01, which is measured at 
the Earth. And these results are in agreement with the result of \cite{Hewitt2012}.

\subsection{Hadron-dominated model}

The middle-lower panel of Fig. \ref{fig:multi-sed} shows the results of the hadron-dominated model 
fitting using the spectra of a power law with an exponential cutoff for protons and the model 
parameters are given in Table \ref{table:model}. In the modeling, the ratio of the number of relativistic 
electrons to protons at 1 GeV, $K_{ep}$, is assumed to be 0.01, which is in accord with the local 
measured CR abundant. The spectral indices of particles are found to be about 1.9. For electrons, 
a break energy of about 6 GeV, a magnetic field strength of about 72 $\mu$G and a total energy 
above 1 GeV of $W_{\rm e}\approx 3.2 \times 10^{47}~{\rm erg}$ are needed to explain the radio spectrum. 
The cutoff energy of electrons, $E_{e, \rm cut} \approx 6.9$ TeV, is determined by making the 
acceleration timescale of diffusive shock with Borm diffusion equal to the synchrotron radiation loss timescale \citep{Parizot2006,Funk2015}, 
\begin{eqnarray}
E_{e, \rm cut} = 83.7 \left(\dfrac{v_{sh}}{10^3 {\rm km}\ {\rm s}^{-1}}\right) \left(\dfrac{B}{1 {\rm \mu G}}\right)^{-1/2} {\rm TeV}
%\left(\dfrac{B}{1 {\rm \mu G}}\right) \left(\dfrac{E_{e, \rm cut}}{1 {\rm TeV}}\right)^2 \left(\dfrac{v_{sh}}{10^3 {\rm km}\ {\rm s}^{-1}}\right)^{-2} = 7\times10^3.
\label{eq:cutoff}
\end{eqnarray}

The electron synchrotron energy loss time at the break energy 6 GeV is about $4 \times 10^5$ years 
which is about 2 orders of magnitude longer than the age of the remnant \citep[4450 year,][]{Becker2012,Winkler1988}
and the shock velocity of $v_{sh}$ = 700 ${\rm km}$~s$^{-1}$ is used \citep{Katsuda2013}.
%The magnetic field strength $B$ is derived from the fitting of different models. 
The total energy of relativistic protons above 1 GeV is
$W_p \approx 7.5 \times 10^{49} (n/4.0\,\mathrm{cm}^{-3})^{-1}\ \mathrm{erg}$,
which is very close to 10\% of the typical kinetic energy released by a core-collapse 
supernova of $E_k\sim10^{51}$ erg. The cutoff energy of protons is about 0.6 TeV, 
which is close to the result of \cite{Hewitt2012}, but lower than that of electrons.

In addition to the spectra of a power law with an exponential cutoff for protons, 
we considered a more complex model: a smoothly broken power law with an exponential 
cutoff, whose form is the same as given by Equation \ref{eq:e_spectra} and the fitted multi-wavelength
SED is compiled in the bottom panel of Fig. \ref{fig:multi-sed}. The break energy of protons, 
$E_{p, \rm br}$, is fitted to be 0.2 TeV to explain the GeV break of SED. And the cutoff
energy of protons, $E_{p, \rm cut}$, should be larger than that of the electrons, 
$E_{e, \rm cut} \approx 6.9$ TeV. Other parameters listed in Table \ref{table:model} 
are the same as that of the power law distribution for protons.

The SBPL model is very similar to the PL model except for a slightly lower TeV fluxes 
of the PL model with the exponential cutoff of the proton distribution.
However the higher cutoff energy of the proton distribution is more reasonable 
and future TeV observations with better sensitivity will be able 
to distinguish these two models.

The hadron-dominated model can naturally explain the radio and GeV spectral
break. Meanwhile, this model expects a not-too-low non-thermal X-ray emission, which is 
more reasonable for Puppis A compared with other young SNRs. Together with the morphology 
of $\gamma$-ray emission, the hadron-dominated model is the most suitable scenario to 
explain the multi-wavelength data of Puppis A.

The above models for the SED show that $\gamma$-ray emission from Puppis A cannot be 
due to interactions of background CRs with intersteller gas.  
In the region of Puppis A the spectral index of CRs inferred from the Galactic diffuse 
$\gamma$-ray emission is about 2.7 \citep{Acero2016,Yang2016}, which is close to the locally observed one 
but much softer than that inferred for Puppis A (see Table \ref{table:model}). 
Furthermore, the mean energy density of CRs responsible for the $\gamma$-ray emission 
of Puppis A, $\varepsilon_p$, derived with $n_{\rm gas}\times W_p /(\dfrac{M_{gas}}{m_p})$,
is higher than that of the background CRs. Therefore, the $\gamma$-ray emission is 
most likely produced by particles accelerated by shocks of the SNR Puppis A. 

\section{Conclusion}

In this work, we re-analyze the $\gamma$-ray emission from SNR Puppis A using seven years Fermi-LAT
data with the latest Pass 8 version. The $\gamma$-ray morphology of Puppis A is well correlated with 
its thermal X-ray and IR emissions rather than the radio emission. This means that some dense clouds may 
exist around Puppis A for providing the target particles of the high energy $\gamma$-ray emission. 
The $\gamma$-ray spectra of Puppis A exhibits a break at $7.92\pm1.91$ GeV with a significance of 
$\sim5.6\sigma$. And the photon indices below and above the break energy are $1.81\pm0.08$ and 
$2.53\pm0.12$, respectively. The existence of the GeV break can naturally explain the upper limit 
derived from HESS observations in the TeV band.

We use three radiation mechanisms, including the ICS-, bremsstrahlung- and hadron-dominated 
models, to fit the multi-wavelength data. Although the ICS- and bremsstrahlung-dominated
models can marginally explain the $\gamma$-ray emission, the hadron-dominated model seems to be
more plausible considering the spectral structure of the GeV break and the radio break detected 
by {\em WMAP} and {\em Planck} satellites. 
However, the decisive evidence of the hadron-dominated model may rely on the detection of 
the characteristic ``$\pi^0$ bump'' in the lower energy band.
Meanwhile, the hadron-dominated model also predicts a not-too-low non-thermal X-ray emission 
of Puppis A, which may be confirmed by the future X-ray observations. 
%Furthermore, the SBPL model suggests that the TeV $\gamma$-ray emission 
%of Puppis A is expected to be detected by the Cherenkov Telescope Array (CTA) in the future.

\section*{Acknowledgments}
This work was supported in part by National Key Program for Research and
Development (2016YFA0400200), 973 Programme of China under grants 2013CB837000 
and 2014CB845800, the 100 Talents program of Chinese Academy of Sciences,
National Natural Science of China under grants 11433009 and 11233001, the Strategic 
Priority Research Program, the Emergence of Cosmological Structures, of the Chinese 
Academy of Sciences, Grant No. XDB09000000.

\end{document}